\documentclass[12pt]{article}
\begin{document}
\baselineskip=18.6pt plus 0.2pt minus 0.1pt
\def\be{\begin{equation}}
\def\ee{\end{equation}}
\def\bea{\begin{eqnarray}}
\def\eea{\end{eqnarray}}
\def\nn{\nonumber\\ }
\def\RR{\mathbb{R}}
\newcommand{\nc}{\newcommand}
\nc{\al}{\alpha} \nc{\bib}{\bibitem} \nc{\cp}{\C{\bf P}}
\nc{\la}{\lambda}
\nc{\C}{\mbox{\hspace{1.24mm}\rule{0.2mm}{2.5mm}\hspace{-2.7mm}
C}}
\nc{\R}{\mbox{\hspace{.04mm}\rule{0.2mm}{2.8mm}\hspace{-1.5mm} R}}
\begin{titlepage}
\title{
\begin{flushright}
{\normalsize \small IFT-UAM/CSIC-05-24 }
\\[1cm]
\mbox{}
\end{flushright}
{\bf A New Mechanism of K\"ahler Moduli Stabilization in Type IIB Theory }\\[.3cm]
\author{Mar\'{\i}a Pilar Garc\'{\i}a del Moral$^{a}$\thanks{{\tt M.G.d.Moral@damtp.cam.ac.uk}}\\[.3cm] {\it
\small $^{a}$ Instituto de F\'{\i}sica Te\'orica, C-XVI, Universidad
Aut\'onoma de Madrid } \\ {\it \small Cantoblanco, E-28049-Madrid,
Spain}\\ 
\\[.2cm]
} } \maketitle
\thispagestyle{empty}

\abstract{We study the scalar potential in supersymmetric (orientifolded) Calabi Yau flux compactifications of
Type IIB theory. We present a new mechanism to stabilize all closed
string moduli at leading order in $\alpha^{'}$ by consistently
introducing fluxes. As usual we consider the dilaton and the complex structure moduli stabilized by turning on three-form fluxes that couple to the F-part of the scalar potential. The Kahler moduli get fixed by the combined action of the flux-induced scalar masses with the magnetic fields of the open string sector, and Fayet-Illiopoulos terms. For supersymmetric three-form fluxes  the model is $N=1$, otherwise the mass terms are the scalar soft breaking terms of the MSSM fields. For the case of imaginary self dual three-form fluxes (ISD), the mass terms are positive and the minimum of the potential is at exactly zero energy. We argue that, under generic assumptions, this is a general
mechanism for the full stabilization of closed string moduli. The vacua
depend explicitly on the
fluxes introduced in the manifold. A
concrete realization of this mechanism for type IIB on a $(\frac{T^{6}}{Z_{2}\times Z_{2}})$ orientifold is provided.}

Keywords: Strings, Branes, Moduli, Phenomenology 
\end{titlepage}
\makeatletter \@addtoreset{equation}{section} \makeatother
\renewcommand{\theequation}{\thesection.\arabic{equation}}


\setcounter{page}{1} \pagestyle{plain}
\renewcommand{\thefootnote}{\arabic{footnote}}
\setcounter{footnote}{0}


\section{Introduction}

The problem of stabilization of moduli fields in string theory (scalar fields with flat
directions in the potential) has been extensively studied, see for example
\cite{ kachru2,gktt,gkp,kklt,kklmmt,kachru-douglas,quevedo1,lust,curio,marchesano1,kachru1,mirjam1,nilles,marchesano2}, since it has
important theoretical and phenomenological consequences. From the
experimental point of view the existence of such massless fields should
be observed in fifth force experiments \footnote{I would like to thank
J. Conlon for this remark.}, but since it is not, the consistency
of the theory requires such fields to acquire a mass. Cosmological
observations also impose constraints on the moduli fixing scale to
reproduce reheating at the inflationary epoch, having as a lower bound
100 TeV. The moduli fields come from two different sectors: closed and open
string sectors. Moduli associated to the closed string sector give
information about the size (K\"ahler moduli) and the shape (complex
structure moduli) of the compact manifold,
and also the dilaton. Moduli associated to the open string sector
correspond to the presence of Wilson lines and to the
parametrization of the position of D-branes in those cases when they are present, in the transverse
dimensions.

A relevant result found in \cite{gkp} showed that a
linear combination of RR, and NSNS three-form fluxes on IIB theory
were able to stabilize the dilaton and the complex structure
moduli. This mechanism presented a great advance in the resolution
of the problem although the K\"ahler moduli remained unfixed. The basic
reason for this is due to the superpotential. It does not contain
any dependence on the K\"ahler moduli, leading to a no-scale scalar
potential at leading order in $\alpha^{'}$. In \cite{kklt} KKLT
found a way to fix one overall K\"ahler modulus by using
non perturbative mechanisms such as condensation of gauginos or
instanton effects. This led to an Anti De-Sitter (AdS) vacuum in a
particular compactification manifold. By explicitly breaking supersymmetry
through the introduction of anti-D3 branes and by fine
tuning fluxes, they were able to lift it to a de Sitter vacuum.
de Sitter vacua have acquired great importance due to the recent
data that suggest the acceleration of the universe and also
because of their close relationship with the inflationary scenario
\cite{kklmmt}.

New
advances in the context of the landscape have been achieved by Douglas et al. \cite{kachru-douglas}
obtaining manifolds with all moduli fixed through  non-perturbative mechanisms and able to lead to
a static cosmology. These last advances, although significant, have not been
able to provide for
realistic compactification manifolds. A potential problem in generating non-perturbative superpotentials from strong infrared dynamics is that it is model dependent. It can also generate too much massless charged matter.

In \cite{quevedo} the model of KKLT was improved. They induce a
supersymmetric model
where one K\"ahler modulus is present by introducing magnetic fluxes
contained on $D7$ branes.

Model building in this context has several fine-tuning and
stability issues. In particular, the superpotential induced by the
fluxes must be hierarchically small $(< 10^{-4})$ in order to
obtain solutions with large volume in which the effective field
theory approximation can be trusted. On the other hand the fluxes must fix the
dilaton at small string coupling to suppress loop effects. Very
recently some papers have appeared \cite{quevedo1}, see also \cite{quevedo2}, that find a way
to stabilize all moduli by considering the combination of
$\alpha^{'3}$ effects and non perturbative contributions to the
superpotential, giving rise to a large volume AdS vacuum.
Its minimum is independent of the
value of the flux superpotential. In this model supersymmetry is broken by
the K\"ahler moduli  and the gravitino mass is not flux
dependent through the superpotential.

Here we propose a  perturbative  mechanism of moduli fixing which is
fullfilled  at the supersymmetric minimum of the theory and it is
able to lead to realistic descriptions with all of the
moduli fixed. In principle one could
think that it is possible
to induce any other dependence in the superpotential
by introducing branes in
the model. However it has been conjectured by \cite{douglas-moore2,douglas-moore1} that B-type
branes (D-branes which wrap 2n-cycles with magnetic fluxes and
the type of branes interesting in IIB models) can couple to the
K\"ahler moduli only through Fayet-Illiopoulos (FI) terms.
Several other works have also studied this
problem. We will consider a IIB theory
compactified on Calabi-Yau (CY) orientifolds, with RR, NSNS 3-form fluxes and
magnetic fluxes. Three-form fluxes couple to the three cycles via the
superpotential associated to the F-term, which does not depend on the
K\"ahler moduli. We show that taking into account the coupling of the
3-form fluxes to the open string sector of magnetized D-branes which
leads to flux induced mass terms, i.e.$\mu$-terms or soft
breaking terms with magnetic fields, together with FI terms
gives a  scalar potential which stabilizes the K\"ahler moduli. This mechanism is model
independent and  we think that it is a generic procedure
for stabilizing moduli in a manifold.
One advantage of this method is that by being supersymmetric or
breaking spontaneously supersymmetry, it can be put
in the 4D standard supergravity form and it keeps control over the types
of interactions that can be induced. In a previous paper \cite{throat,thesis} a method was proposed to
stabilize some K\"ahler moduli through the coupling of fluxes to the
FI-term in order to fix  the blow-up moduli of the model. In
that case the expected soft breaking term contribution did not
include magnetic fields and failed in the attempt of stabilizing the full K\"ahler moduli. 
This idea is an extension of that one including magnetic fields in the
configurations. In \cite{graham}, in a different approach, they also
consider mass terms as a mechanism to stabilize moduli.

In \cite{antoniadis,antoniadis2} a type I theory was proposed with three-form fluxes
and magnetic fluxes. They claim that they are able to stabilize K\"ahler moduli just
through FI-terms. We argue that what they find does not constitute a true stabilization
of the moduli since the moduli are free to acquire any other vev without any energy cost. 
To stabilize K\"ahler moduli it is necessary to also have a coupling of fluxes to the open
 string sector.

The paper is organized as follows: in section 2 we give a brief summary
of three-form flux stabilization. We show how a linear combination of three
form fluxes stabilizes complex structure moduli and the dilaton under
very general assumptions. Non perturbative mechanisms are used to
stabilize K\"ahler moduli. In section 3 we review how soft
terms appear. We par\-ti\-cu\-lar\-ly focus on the flux induced soft breaking terms
with magnetic fields and we find a general expression for these
terms in a toroidal orientifold generalizing to the case when all K\"ahler moduli are different. In section 4 we describe D-term
supersymmetry breaking. FI terms that couple to B-type branes represent
a deviation from the supersymmetry conditions for the branes and a shift in
the value at which the moduli have a supersymmetric value. However a proper recombination
mechanism can restore the supersymmetry. In section 5 we
propose the mechanism for stabilizing moduli without introducing non perturbative 
mechanisms and we describe in detail the minimization of the scalar
potential. In section 6 we provide a concrete realization of this
mechanism (of phenomenological interest) for the case of ISD three-form fluxes. We perform
IIB compactified on $T^{6}/Z_{2}\times Z_{2}\times \Omega R$ moduli stabilization and we
indicate the explicit values at
which K\"ahler moduli get fixed. We conclude in section 7 with a discussion, summarizing the main results.

\section{Review of moduli stabilization}

In type IIB theory on a Calabi-Yau manifold the closed string moduli content associated to the geometry are: an axion-dilaton $S$, $h_{11}$ K\"haler moduli $T_{i}$ parametrizing the $CY_{3}$ size  and $h_{21}$ complex structure moduli $U_{i}$ parametrizing the $CY_{3}$ shape, where $h_{11}, h_{21}$ are the Hodge numbers characterizing the Calabi-Yau three-fold.
The K\"ahler potential associated to the closed string sector has,
for toroidal compactifications where the metric factors are into three two-by-two blocks, the following  expression,
\bea
k^{2}\mathcal{K}(S,U_{i},T_{i})=-ln[S+\overline{S}]-\sum_{j=1}^{h_{11}}ln[T_{j}+\overline{T}_{j}]-\sum_{j=1}^{h_{21}}ln[U_{j}+\overline{U}_{j}].
\eea

Giddings, Kachru and Polchinski \cite{gkp} introduced the methods to stabilize the dilaton
and complex structure moduli by turning on 3-form fluxes.
In type IIB theory,
strings can have RR and NSNS 3-form field strengths, which
can wrap dual 3-cycles labeled by A and B leading to quantized background
fluxes,
\bea
\frac{1}{4\pi^{2}\alpha^{'}}\int_{A}F_{3}=M\quad
\frac{1}{4\pi^{2}\alpha^{'}}\int_{B}H_{3}=-K
\eea
where K and M are arbitrary integers. These forms allow consistent
string  compactifications of generic orientifold $CY_{3}$ manifolds. Fluxes
also have some other interesting consequences: they induce a warp
factor in the metric that deforms the manifold and generates hierarchies \cite{throat},
fix the moduli partially, and also give a mechanism
to break supersymmetry in a controlled manner by inducing soft breaking
terms.

To understand the mechanism of the partial fixing of moduli we have to remark that these fluxes generate a superpotential found by \cite{gukov}
\bea
W=\int_{CY_{3}}G_{3}\wedge\Omega_{3},
\eea
where $G_{3}=F_{3}-SH_{3}$, with $S$ the complex axion-dilaton of type
IIB theory. $\Omega_{3}$ is the unique (3,0) form of
the Calabi-Yau threefold. The holomorphic three form $\Omega_{3}$ has a non-trivial dependence on the complex
structure moduli $U_{i}$. This superpotential is independent of the K\"ahler moduli $T_{i}$. It gives the $D=4, N=1$ scalar
potential
\bea
V_{F}=\exp^{k^{2}\mathcal{K}}(K^{I\overline{J}}D_{I}W\overline{D_{J}W}-3k^{2}\vert W\vert^{2})
\eea
Here $I,J$ label all the above geometric moduli of the manifold.

The covariant derivatives are defined as
$D_{I}W=\partial_{I}W+k^{2}\partial_{I}\mathcal{K}W$ where $\mathcal{K}$
denotes the K\"ahler potential and $K^{I\overline{J}}$ the inverse of the K\"ahler metric defined in terms of the K\"ahler potential $K_{I\overline{J}}=\partial_{I}\partial_{\overline{J}}\mathcal{K}$.

This potential is of no-scale type ($\Lambda=0$ at tree level in $\alpha^{'}$)
since the K\"ahler dependence on $T_{i}$ cancels exactly the $3W^{2}$
contribution. Since the potential is positive definite, the global minimum of this potential lies at zero. The values of complex structure moduli and the dilaton
get fixed for the particular values at which the superpotential is
minimized, $D_{i}W=0$, where $i$ runs over all fields  except the K\"ahler moduli. The K\"ahler moduli however remain unfixed as the
superpotential has no dependence on them. The minimum of the potential is supersymmetric if $D_{T}W=\partial_{I}\mathcal{K}W=0$ and non-susy otherwise.

The K\"ahler moduli have been stabilized by a different mechanism, namely non-perturbative contributions that introduce an exponential dependence.
This contribution together with the one induced by  fluxes gives the total superpotential
\bea
W=W_{flux}+A \exp(-aT)
\eea
and  generate a scalar potential which typically has an AdS minimum at a finite value of
$T$ and a run-away behaviour at infinity. Several mechanisms for
breaking supersymmetry allow in principle this minimum to be lifted to
a de Sitter vacuum, i.e. \cite{kklt,quevedo,quevedo3}.

\section{Soft breaking terms with magnetic fields}
The low energy effective action can have susy-breaking soft terms
coming from magnetized or non magnetized configurations. Soft terms
are operators of $dim<4$ which do not induce quadratic divergences
that spoil the good properties of the $N=1$ supersymmetry. They give
mass to the superpartners of the SM fields, when spontaneous
supersymmetry breaking occurs. Their scale is expected to be not much
above the electroweak scale. In fact when they are induced through
fluxes they are of the order of the flux scale. These soft terms arise
from the interaction of low dimensional D-brane charges induced on the
D7 branes with the background flux (see \cite{camara2,camara1}). The
soft terms contain gaugino masses, trilinear terms, and scalar masses
of MSSM fields. Regarding the stabilization of K\"ahler moduli we are
only going to be interested in scalar masses since they are the
dominant terms in the expression.
In this section we want to extend the results found in \cite{anamaria}, (see also \cite{reffert1,reffert2}) to more general settings. Unless specified otherwise we will follow the notation of \cite{anamaria} but extending their results to be valid for a general configuration of D-branes compactified on toroidal orientifolds of type IIB theory on $T^{6}/Z_{2}\times Z_{2}\times \Omega R$ .

We consider a six torus factorized as
$T^{6}=\otimes_{i=1}^{3}T^{2}_{i}$, performing the quotient by
$Z_{2}\times Z_{2}\times \Omega R$ symmetry as in \cite{marchesano1},
filled with D7 branes some of which are magnetized and D9 branes
carrying anti-D3 brane charges in the hidden sector to cancel RR
and NSNS tadpole conditions while preserving $N=1$ supersymmetry.
The magnetic constant field is defined in terms of the wrapping number $m$ of each stack of $D7_{a}$ branes transverse to the torus $i$, and $n$ which represents the units of magnetic flux.
\bea
\frac{m_{a}^{i}}{2\pi}\int_{T^{2}_{i}}F^{i}_{a}=n^{i}_{a}
\eea
$F^{i}_{a}$ is the world-volume magnetic field. For later convenience we define the following angles,
\bea
\Psi^{i}_{a}=arctan (2\pi\alpha^{'}F^{i}_{a})=arctan(\frac{\alpha^{'}n^{i}_{a}}{m^{i}_{a}A_{i}})
\eea
$A_{i}$ represents the area of the two torus.
Open strings give rise to charged fields. There are two types of
states living on stacks of Dp-branes. {\tt Untwisted} states are
chiral fields coming from open strings whose ends are attached to
the same stack of branes and {\tt twisted} sector chiral fields those
lying at two different stacks of magnetized Dp-branes. We
are only going to consider the twisted sector fields as they are the
ones that have bosonic soft breaking terms. From D7 brane twisted
sectors $D7_{i}-D7_{j}$ there are only chiral massless multiplets
denoted by $C^{7{i}7_{j}}$ transforming in the bifundamentals of
$G_{i}\times G_{j}$, with $G_{i}$ being the gauge group associated
to the enhanced symmetry of each stack of D7 branes. The low
energy dynamics of the massless fields is governed by a $D=4, N=1$
supergravity action that depends on the K\"ahler potential, the
gauge kinetic functions and the superpotential. The moduli in
this type of constructions, as already explained in
\cite{anamaria}, are $\{M,C_{I}\}$ with $M$ the closed string moduli
and $C_{I}=\{C^{7_{i}}, C^{7_{i}7_{j}}\}$, where $C^{7_{i}}_{i}$
are the moduli fields representing the position of the $D7_{i}$
branes in the transverse $T_{i}^{2}$ whereas $C_{j}^{7_{i}}$
correspond to the presence of Wilson lines turned on, in the two
complex directions parallel to the $D7_{i}$ brane. In the
following we will not care about open string moduli since there
are models of phenomenological interest free of them, such as the
one we propose in the last section. Closed geometric moduli that
appear in the 4D $N=1$ supergravity action consist of the
complex dilaton \bea S=e^{-\phi_{10}}+ia_{0} \eea where $a_{0}$
is the R-R 0-form and $e^{\phi_{10}}=g_{s}$ the string
coupling constant; the complex structure moduli $U_{j}, j=1,2,3$,
which are equivalent to the following geometrical moduli for the
case of toroidal compactifications, \bea
\tau_{j}=\frac{1}{e_{jx}^{2}}(A_{j}+i e_{jx}. e_{jy}) \eea
where $e_{jx},e_{jy}$ are the $T^{2}_{j}$ lattice vectors, $A_{j}$ is the
area of the two-torus, which for the particular case of the square
$T^{2}_{j}$ is equal to $A_{j}=R_{jx}R_{jy}$, and the dual angle
is
$\varphi_{a}^{i}=\arctan(\frac{n_{a}^{i}R_{ix}}{m_{a}^{i}R_{iy}})$.  The geometric K\"ahler 
moduli  $\rho_{i}$ are described by,
\bea
\rho_{j}=
A_{j}+ia_{j},\eea where the axions $a_{j}$ arise from the R-R 4-form. However, as explained in \cite{anamaria,reffert1,reffert2} 
 the K\"ahler moduli field denoted
by $T_{i}, i=1,2,3$ are not equivalent to the
geometric moduli $\rho_{i}$,
since its correct expression is\bea 
T_{i}=\exp^{-\phi_{10}}\frac{A_{j}A_{k}}{\alpha^{'2}}+ia_{i}\quad
j\ne i\ne k \eea and the gravitational coupling in $D=4$ is
$G_{N}=k^{2}/8\pi$ with \bea k^{-2}=\frac{M^{2}_{pl}}{8\pi}
=\exp^{-2\phi_{10}}\frac{A_{1}A_{2}A_{3}}{\pi\alpha^{'4}}. \eea The
string scale is defined as $M_{s}=\frac{1}{\sqrt{\alpha^{'}}}$.
We have chosen this type of compactification for its simplicity:
The moduli in this type of compactification are just the dilaton
$S$, three complex structure moduli associated to the relation
between the radius of each tori $U_{i}, i=1,.,3$ and three K\"ahler
moduli associated to the size of each torus $T_{i}, i=1,2,3$.
\\
The K\"ahler potential contains the K\"ahler part $\widehat{K}$
associated to the closed string moduli, $M=\{S,T_{i},U_{i}\}$, and the K\"ahler part
$\widetilde{K}$ which is associated to the matter fields
$C_{I}=\Phi_{aa},\Phi_{ab}$ of the open string sector
\cite{anamaria}. 
\bea
k^{2}\mathcal{K}&=&k^{2}\widehat{K}+k^{2}\sum_{IJ}\widetilde{K}_{IJ}C_{\overline{I}}C_{J}+k^{2}\sum_{IJ}Z_{IJ}(C_{I}C_{J}+c.c)+\dots
\eea

The standard expression for the soft breaking terms found in
\cite{brignole} is \bea m_{I}^{2}=m_{3/2}^{2}+
V_{0}-\sum_{M,N}\overline{F}^{\overline{M}}F^{N}\partial
_{\overline{M}}\partial_{N}log(\widetilde{K_{IJ}}) \eea
in supergravity conventions with all quantities measured in Planck
units. $I=aa,ab,ba, bb$ labels the different stacks of magnetized D7 branes.
$F^{M}$ are the auxiliary fields of the corresponding chiral
multiplet $\Phi_{A}$ which in general have the following
expression \bea
\overline{F}^{\overline{A}}=k^{2}\exp^{k^{2}K/2}K^{\overline{A}B}D_{B}W
\eea where $\Phi_{A}=\{M,C_{I}\}$ and as before
$C_{I}=\{C_{j}^{7_{i}},C^{7_{i}7_{j}}\}$. When a spontaneous breaking of supersymmetry occurs, the auxiliary vevs acquire a vacuum expectation value. In this case the supersymmetry breaking is produced by the presence of a non supersymmetric 3-form flux, ($G_{3}$ containing a $(0,3)$ piece).  The auxiliary fields
are parametrized as
\bea F^{S}=\sqrt{3}Csm_{3/2}\sin(\theta)\exp^{-i\gamma_{s}},\\
F^{T_{i}}=\sqrt{3}C
t_{i}\eta_{i}m_{3/2}\cos(\theta)\exp^{-i\gamma_{i}}, \eea such that
$\sum_{i}\eta_{i}^{2}=1$ and $\eta$ and $\theta$, the goldstino
angle, control whether $S$ or $T_{i}$ dominate the SUSY breaking.
Here \bea C^{2}&=&1+\frac{V_{0}}{3m_{3/2}^{2}}\\
m_{3/2}&=&\exp^{1/2k^{2}\mathcal{K}}W\\
V_{0}&=&F^{m}\widehat{K}_{mn}F^{n}-3m^{2}_{3/2} \eea and $m_{3/2}$
denotes the gravitino mass and $V_{0}$ the cosmological constant.

For the case of soft breaking terms induced by 3-form and
magnetic fluxes, the expression becomes \bea m^{2}_{I}&=
m^{2}_{3/2}+V_{0}-1/4\overline{F}^{\overline
M}F^{N}\partial_{\overline{M}}\partial_{N}\ln(st_{1}t_{2}t_{3}) +\sum_{i=1}^{3}(\partial_{\overline{M}}\partial_{N}
\nu_{i})\overline{F}^{\overline
M}F^{N}(ln(u_{i}))\\ \nonumber &-1/2\sum_{i=1}^{3}\overline{F}^{\overline
M}F^{N}\partial_{\overline{M}}\partial_{N}(\ln(\Gamma(1-\nu_{i}))-\ln(\Gamma(\nu_{i}))).
\eea 
where the K\"ahler potential has the following form, \bea
k^{2}\mathcal{K}&=&k^{2}\widehat{K}+k^{2}\sum_{IJ}\widetilde{K}_{IJ}C_{\overline{I}}C_{J}+k^{2}\sum_{IJ}Z_{IJ}(C_{I}C_{J}+c.c)+\dots\\
\widetilde{K}&=&\frac{(st_{1}t_{2}t_{3})^{\frac{1}{4}}}{2\pi\alpha^{'}}
\Pi_{j=1}^{3}u_{j}^{-\nu_{j}}\sqrt{\frac{\Gamma(1-\nu_{j})}{\Gamma(\nu_{j})}}\\
k^{2}\widehat{K}&=&-\ln(s)-\sum_{i}\ln{t_{i}}-\sum_{i}\ln(u_{i})
\eea with 
the conventions of \cite{anamaria} \bea
s=S+\overline{S};\quad t_{i}=T_{i}+\overline{T_{i}}; \quad
u_{i}=U_{i}+\overline{U_{i}}. \eea 
$k^{2}=4\pi\alpha^{'}(st_{1}t_{2}t_{3})^{-1/4}$, and
$\widehat{\nu}_{i}=\frac{1}{\pi}(\Psi^{i}_{ab})=\frac{1}{\pi}(\Psi^{i}_{b}-\Psi^{i}_{a})$,
where $\Psi^{i}_{a}=\arctan(g_{a}\beta_{i})$ with
$\beta_{i}=\sqrt{\frac{st_{i}}{t_{jt_{k}}}}$ and
$g_{a}^{i}=\frac{n_{a}^{i}}{m_{a}^{i}}$. 
$\sum\widehat{\nu}_{i}=0$ trivially, which is the condition for
two stack of D-branes to preserve the same supersymmetry.
The $\nu_{i}$ are computed in terms of
$\widehat{\nu}_{i}/\pi$, such that $0<\nu_{i}<1$ and
$\sum_{i=1} \nu_{i}=2$ as in \cite{anamaria}. Here 
$\nu_{i}=1+\widehat{\nu}_{i}/\pi$ iff $\widehat{\nu}_{i}\le 0$ and
$\nu_{i}=\widehat{\nu}_{i}/\pi$ otherwise.
 This last expression is closely related to the one in   \cite{anamaria}, although it has been generalized to the case when all of the K\"ahler moduli are different.

Regrouping terms, \bea
m^{2}_{ab}&=m^{2}_{3/2}+V_{0}-1/4\overline{F}^{\overline
M}F^{N}\partial_{\overline{M}}\partial_{N}\ln(st_{1}t_{2}t_{3})+
\sum_{i=1}^{3}(\partial_{\overline{M}}\partial_{N}
\nu_{i})\overline{F}^{\overline
M}F^{N}\\ \nonumber
&(ln(u_{i})-1/2B_{0}^{i}(\nu_{i}))
-1/2\sum_{i=1}^{3}B_{1}^{i}(\nu_{i})\overline{F}^{\overline
M}F^{N}(\partial_{\overline{M}}\nu_{i}\partial_{N}
\nu_{i}).
\eea 
The $B_{p}$ are defined in terms of polygamma functions as, \bea
B_{0}^{i}(\nu_{i})&=&\frac{\partial_{\nu}\Gamma(1-\nu_{i})}{\Gamma(1-\nu_{i})}-
\frac{\partial_{\nu}\Gamma(\nu_{i})}{\Gamma(\nu_{i})},\\
B_{p}^{i}&=&\partial_{\nu}B_{p-1}(\nu_{i}),\eea and in terms of a
useful analytical expression is \bea B_{0}(z)=\pi cotan(\pi z)
\eea and its derivatives. Let us remark that this definition of
$B_{p}(\nu)^{i}$ is slightly different of the one used in
\cite{anamaria}. A special value of this function is at
$\nu_{i}=1/2$ where $B_{k}^{i}(1/2)=0$.

We finally obtain the following expression, \bea
m^{2}_{ab}&=&m^{2}_{3/2}+V_{0}-3/4C^{2}m^{2}_{3/2}[1
+1/4\pi^{2}\sum_{i=1}^{3}(ln(u_{i})-1/2 B_{0}^{i}(\nu_{i}))\\
\nonumber &&
{}[4\mathbf{P}_{i}\sin(2\pi\Psi_{i})_{ab}-\mathbf{Q}_{i} \sin(4\pi\Psi_{i})_{ab}]
+1/8\pi^{3}\sum_{i}^{3}B_{1}^{i}(\nu_{i})\mathbf{Q}_{i}(\sin
(2\pi\Psi_{i})_{ab})^{2}]\eea where
$\sin(\pi\Psi_{i})_{ab}=\sin(\pi\Psi_{i})_{b}-\sin(\pi\Psi_{i})_{a}$.
and the above variables $\mathbf{P},\mathbf{Q}$ are defined in
terms of goldstino angles, \bea \mathbf{P}_{i}&=&
\sin^{2}\theta+\cos^{2}\theta(\eta_{i}^{2}-\eta_{j}^{2}-\eta_{k}^{2})\\
\mathbf{Q}_{i}&=&1- 2\cos^{2}\theta \{
\eta_{i}\eta_{j}cos(\gamma_{i}-\gamma_{j})+\eta_{i}\eta_{k}cos(\gamma_{i}-\gamma_{k})-
\eta_{j}\eta_{k}cos(\gamma_{j}-\gamma_{k})\}\\ \nonumber
&& {}+\sin(2\theta)\{ \eta_{i}cos(\gamma_{i}-\gamma_{s})-
\eta_{j}cos(\gamma_{j}-\gamma_{s})-\eta_{k}cos(\gamma_{k}-\gamma_{s}\}
\eea so they are sensitive to the particular choice of 3-form
fluxes.

The expression for soft breaking mass terms can be finally expressed as,
\bea
m^{2}_{ab}=\frac{1}{4}(1+3A)m^{2}_{3/2}+\frac{1}{4}(3+A)V_{0} \eea
where \bea A&\equiv&-1/4\pi^{2}\sum_{i=1}^{3}(ln(u_{i})-1/2
B_{0}(\nu_{i}))
{}[4\mathbf{P}_{i}\sin(2\pi\Psi_{i})_{ab}-\mathbf{Q}_{i}
\sin(4\pi\Psi_{i})_{ab}]\\
&&-1/8\pi^{3}\sum_{i}^{3}B_{1}(\nu_{i})\mathbf{Q}_{i}(\sin
(2\pi\Psi_{i})_{ab})^{2}. \eea

These results recover the ones of \cite{anamaria} by making appropriate
substitutions for their particular configuration and imposing
$t_{2}=t_{3}$.
Although the K\"ahler part of the potential
$\widetilde{K}_{C\overline{C}}$ correspond to the twisted magnetized
sectors, with the above definitions, the soft terms at twisted, unmagnetized
$D7_{2}-D7_{3}$ stacks are also correctly found due to appropriate cancellations.

\section{ D-term supersymmetry breaking}
In this section we are interested in characterizing D-term behaviour in the presence of B-type branes. The motivation is the following: in order to solve the problem of K\"ahler moduli stabilization, naively we can think of finding a perturbative generalization of the superpotential that contains K\"ahler moduli in its definition. However this seems not to be possible. K\"ahler moduli are the moduli associated to the $(1,1)$ forms that naturally can be thought of as being stabilized by magnetic fluxes living on D-branes. In general there are $h_{11}$ of them, being the $h_{11}$ Hodge number counting the number of 2-cycles present in the compactified manifold. The Decoupling Statement of Douglas e al.
\cite{douglas-moore2,douglas-moore1} establishes that K\"ahler moduli
on Type IIB can only couple to B-type branes (i.e. Branes wrapping
even cycles) through Fayet Iliopoulos (FI) terms in the scalar potential. The mirror
statement of this on type IIA says that complex structure moduli can only couple
to A-branes (branes wrapping odd cycles (3-cycles)) through FI terms. This fact
seems disappointing from the perspective of finding a perturbative
superpotential for K\"ahler moduli. Moreover this is a common
feature for the case of D-branes at singularities \cite{douglas-moore2, douglas-moore1}.

In \cite{cvetic1, cvetic*} an
$\mathcal{N}=1$ 4D type IIA theory on $\frac{T^{6}}{Z_{2}\times Z_{2}}$
orientifold with D6 branes at angles was studied. In this section we
will review the main properties. This model is dual to magnetized D9
branes on type IIB theory with discrete torsion. Requiring
supersymmetric models imposes a condition between orientifold planes and D branes. On type IIA, in a supersymmetric model
each stack of D6 branes is related to the orientifold planes O6 by
a rotation in $SU(3)$. The supersymmetry configurations imposes a
condition on the angles $\theta_{i}$ of the D6 branes with respect to the
orientifold plane in the i-th direction of the two-torus, which for
the case they considered was, \bea \sum_{i=1}^{3}\theta_{i}=0. \eea This
condition is equivalent to \bea
\sum_{i=1}^{3}\arctan{(\chi_{i}\frac{m_{i}}{n_{i}})}=0, \eea
with $R_{xi},R_{yi}$ the radius of the i-th
two-torus and $\chi_{i}=\frac{R_{yi}}{R_{xi}}$ the untwisted complex structure moduli.
 This model is T-dual to a type IIB orientifolded theory with discrete
torsion and  twisted K\"ahler moduli with the condition
\bea \sum_{i}\Psi^{i}_{a}&=&\frac{3\pi}{2}mod(2\pi),\eea
where we are using the convention for angles introduced in the preceding section, $\Psi^{i}_{a}=arctan(\frac{\alpha^{'}n^{i}_{a}}{m^{i}_{a}A_{i}})$ in terms of the tori areas $A_{i}$.
$A_{i}$ are expected to couple open string modes in the D9 branes on
2-cycles (B-branes) as Fayet-Illiopoulos  (FI) term. This is the dual version to the one indicated in
\cite{cvetic1}. In \cite{witten} it is explained that FI terms are proportional to the deviation
from the susy condition, giving an effective action proportional to the deviation,
\bea
\sum_{a}\int dx^{4}\xi_{a}D_{a}
\eea
where for small
FI terms , \cite{luis2}, $\xi_{a}=\delta\sum_{i}\Psi_{a}^{i}=(\sum_{i}\Psi_{a}^{i}- susy\quad cond.)$
and vanishes for supersymmetric configurations of D-branes.
Since
$D_{a}=\sum_{b}^{N_{b}}\left|\phi_{ab}\right|^{2}-\left|\phi_{ba}\right|^{2}+\xi_{a}$, and
$V_{D}=\frac{1}{2}D_{a}D^{a}$ then, the D-term piece  of the scalar potential is then
\bea
V_{D}=\sum_{a}^{N_{a}}(\sum_{b}^{N_{b}}\left|\phi_{ab}\right|^{2}-\sum_{b}\left|\phi_{ba}\right|^{2}+\xi_{a})^{2},\eea
where $\left|\phi_{ab}\right|$ represents the charged matter fields
lying on an oriented string with ends attached at two different
stacks $a$ and $b$. $\left|\phi_{ba}\right|$ represents the matter fields at the
intersection of the same two stacks $a$ and $b$ with reversal
orientation and has also to be considered. $N_{a},N_{b}$ are the
number of parallel Dp-branes at each stack. 

The dependence of the scalar potential on a $\phi_{ab}$ matter field is
\bea
V_{D}(\phi_{ab})=(\left|\phi_{ab}\right|^{2}+1/2\xi_{a})^{2}+(-\left|\phi_{ab}\right|^{2}+1/2\xi_{b})^{2}
\eea
where we have renamed for the second part of the R.H.S $a \to b$ and
we have used that $\left|\phi_{ba}\right|^{2}=-\left|\phi_{ab}\right|^{2}$
and  the mixed term which gives a mass term for $\left|\phi_{ab}\right|$
\bea
\left|\phi_{ab}\right|^{2}(\xi_{a}-\xi_{b})=\left|\phi_{ab}\right|^{2}\delta\Psi_{ab},
\eea
defining $\delta\Psi_{ab}=\xi_{b}-\xi_{a}$ and approaching
$(\delta\Psi_{ab})^{2}\sim\frac{1}{2}(\xi_{a}^{2}+\xi_{b}^{2})$,  we arrive to the
familiar expression for D-terms appearing from twisted sector, \cite{luis2}:
\bea
V_{D}=\sum_{I}(q_{I}\left|C_{I}\right|^{2}-\delta\Psi_{I})^{2}
\eea
where $I$ runs over all the indices of the matter field , i.e,
$aa,ab,ba,bb$. $q_{I}$ are the charges of the matter fields  under the
$U(1)$ gauge group of the D7 branes, with $q_{aa}=0,q_{ab}=-q_{ba}, q_{ab}=+1,0,-1$; and $C_{I}=\phi_{aa},\phi_{ab}$ are all of the open string moduli. The FI terms appear always when there are supersymmetry breaking coming from the 2-form magnetic fluxes.
In standard type IIB orientifolded actions the FI terms can be properly tuned by adjusting the twisted moduli. Physically this term reproduces at
leading order the splitting between scalar and fermion masses \cite{cvetic1}. Namely in the
$ab$ sector, chiral fermions remain massless at tree level
while their scalar partners obtain a mass proportional to $\delta\nu_{ab}=\frac{1}{\pi}\sum_{i}(\Psi_{a}^{i}-\Psi_{b}^{i})$.

An important remark regarding supersymmetry is the following:
In \cite{cvetic1} it has been pointed out that since some of these scalar fields can acquire vevs, the existence of a
FI terms by itself does not automatically imply a susy breaking since
they may acquire them so as to make the D-term vanish. Physically it is due to a
recombination process of the D6 branes -in type IIA picture- (which now are not
supersymmetric since the angles have changed) into supersymmetric
smooth 3-cycles. This process gives a vev to some of the scalar fields
$\phi_{ab}$, at the intersection.

In the type IIB picture, the argument remains valid. D9 branes have magnetic fields at supersymmetric values. In the presence of non-vanishing FI terms the D9 branes become non susy, but if some scalars acquire mass cancelling D-term contribution, then they change their magnetic values on the 2-cycles restoring supersymmetry.

It has also been argued that although the existence of a
FI-term in local compactifications allows for supersymmetry breaking,
for the case of  global compactifications recombination processes seem
to be a generic mechanism to restore supersymmetry.

As already mentioned in the introduction, in \cite{antoniadis,antoniadis2}  a mechanism was proposed to stabilize K\"ahler
moduli through three-form fluxes and magnetic fluxes. The argument was
that the K\"ahler moduli get fixed at its supersymmetric value since the
supersymmetry condition for magnetized D-branes depends explicitly on
them. However, because of the argument above,  this is not a true
stabilization since matter fields can acquire a vev without any energy cost to cancel the D-term
contribution, modifying the value of the twisted moduli. Then the scalar
potential has a flat direction with $V=0$.

\section{ A new mechanism of moduli stabilization}
We are going to consider a decoupling approach where complex
structure moduli and the dilaton have been stabilized by the
standard mechanism. Turning on suitable three-form fluxes stabilizes their vevs and allows their dynamics to be integrated out as
in \cite{kklt}. This approach is valid in the regime when the mass scale of the K\"ahler moduli is much less than that of the dilaton and the complex structure moduli. The warping induced by the fluxes will be negligible in the approximation of large volume.
We
propose the following mechanism to stabilize the K\"ahler moduli
perturbatively: We consider the coupling of fluxes (3-form fluxes and magnetic ones) to the open string sector that induce
bosonic flux induced masses. These mass terms in principle can be soft breaking
terms or  supersymmetric.  Supersymmetric mass terms are the $\mu$ terms appearing in the
superpotential when supersymmetric 3-form fluxes (2,1) are turned on
and the D-brane stacks share at least one parallel direction
\cite{reffert2,angelantonj}.  These flux induced masses $m^{2}_{I}$ for magnetized
D-branes combined with the FI contribution represent a new coupling in
the scalar potential that lift the flat directions of the potential
giving masses to all the moduli. This fact will happen generically
when soft masses are present but as remarked above, for 
supersymmetric masses to exist on susy flux-backgrounds, only very specific configurations will
be allowed. The effective scalar potential in $D=4$
taking into account the F-piece and the D-piece of the potential is equal to
\bea
V_{T}&=&V^{back}_{F}+\textrm{flux induced mass terms}+
FI=\\&=&V_{F}^{back}+\sum_{I}m^{2}_{I}|C_{I}|^{2}+\sum_{I}(q_{I}|C_{I}|^{2}-\sum_{i}\delta\Psi_{I}^{i})^{2},
 \eea 
where $V_{F}^{back}$ is the background no-scale potential induced by
the superpotential. Flux induced mass terms have the effect of lifting
the flat directions in the potential. These terms can be positive or
negative.  For the case of ISD $G_{3}$ the mass terms are positive and
 $V$ is also positive definite. We will see that there exists a global
minimum of the potential $V$ that fixes all of the
moduli. In order to minimize this potential we should consider the
minimization with respect to every moduli $C_{I}, M$. It is important
to ask 
whether the
resulting critical point is a saddle point or a minimum. The
answer is that iff there exist a critical point such that $V_{T}=0$, 
being $V_{T}$ definite positive, then this point is a global minimum
of the theory and in this case it corresponds to its supersymmetric value. (The supersymmetric minimum 
in fact corresponding to F-flatness and D-flatness condition)
since it is bounded from below to a value greater or equal than $0$.
We find the value of this minimum at 
$|C_{I}|=0,\sum_{i}\delta\Psi^{i}_{I}=0$.

 Let see it in more detail. To see the stabilization we need to
minimize the potential with respect to the full moduli $C_{I}, M$,
since the flat directions were associated in \cite{cvetic1} to the
presence of matter fields which did not acquire a vev. As explained before, the non-vanishing of the FI term represents a shifting from the supersymmetric condition associated to the D-brane configuration
$\sum_{i}\delta\Psi_{I}^{i}\ne 0$.
As we are interested in stabilizing K\"ahler moduli on type IIB,
then we are going to impose that the flux induced terms have to
be associated to the presence not only of 3-form fluxes but also of
magnetic fields which depend explicitly on the K\"ahler moduli. The presence of
magnetic fluxes gives an extra dependence of the intersecting
angles $\sum_{i}\delta\Psi^{i}_{I}$ on the K\"ahler moduli.

The superpotential is not renormalized at
any order in perturbation theory and receives no $\alpha^{'}$
corrections from the bulk, however it could receive from brane
contributions \cite{grana}. The D-term, by an appropriate selection of
fluxes, can leave the model supersymmetric and so will not receive corrections
changing its structure.
Moreover,
if we analyse the equation we can see that this holds generically 
given $m^{2}_{I}>0$ (as we will see this is the case for the interesting
ISD 3-form fluxes). 

We have in principle two kinds of flux contributions,
ISD fluxes and IASD fluxes. We do not consider interaction terms
between them. Since IASD fluxes do not solve type IIB 10D equations of
motion unless non-perturbative effects will be taken into account, we will not consider them in the analysis. In the presence of ISD fluxes only, as pointed out
in \cite{camara1}, soft terms are positive. The argument relies in the
following : they can be regarded as
geometric moduli of the F/M-theory fourfold and generate positive
definite scalar potential. In \cite{luis} it is argued that this is a
general property of ISD fluxes also valid in the case of magnetic fluxes at the
intersections \footnote{We thank L. Ibanez for helpful clarifying
explanation at this point}. A way of illustrating this point is that
for the case of soft breaking terms with magnetic fluxes, the
superpotential \bea W=\int_{CY_{4}}G_{4}\wedge \Omega_{4} \eea
leads to a positive definite scalar potential. $W$ in particular
includes D7-brane geometric moduli \cite{W}, which in this case are described in
terms of the homological charges also coming from the magnetized D-branes.

For the case of ISD fluxes, the $V_{F}^{TT}=0$ since it is a no-scale potential ($D_{T}W\overline{D_{T}}W=3\left|W\right|^{2}$), then
extremizing with respect to the moduli $T_{N}$ gives the equation
\bea
\partial_{N}V=0\to \sum_{I}(-2q_{I}|C_{I}|^{2}\partial_{N}\Psi_{I}
+2(\delta\Psi_{I})\partial_{N}\Psi_{i}+(\partial_{N}m^{2}_{I})|C_{I}|^{2})=0.
\eea Minimizing with respect matter field gives as solution \bea
|C_{I}|&=&0,\quad  V(C_{I})=|\delta\Psi_{I}|^{2}\\
|C_{I}|&=&\pm\sqrt{\frac{2q_{I}\delta\Psi_{I}-m^{2}_{I}}{2q_{I}^{2}}},\quad V_{D}(|C_{I}^{2}|_{min})=\frac{m^{2}_{I}}{4q_{I}^{2}}(4q_{I}(\delta\Psi_{I})-m^{2}_{I}).
\eea
The first solution imposes in this model that the global minimum lies
at $\delta\Psi_{I}=0$,
fixing the moduli through the dependence of $\Psi_{I}$ to its
supersymmetric value iff $\textrm{soft terms}>0$ for all $I$. The value of the minimum of the potential corresponds for supersymmetric configuration of D-branes, to a no-scale potential ($V=0$), and in those cases when supersymmetry can be consistently broken through the FI term to a de Sitter minimum, in the same spirit as \cite{quevedo}.
The other two extrema lead to an AdS vacua and they are only possible
in those cases when $\left|C_{I}\right|$ is real, that correspond to
have negative squared mass terms for the supersymmetric case,
$m^{2}_{I}<0$, so it is not possible for ISD $G_{3}$ fluxes or to have
$0<m_{I}^{2}<2\sum_{I}q_{I}\delta\Psi_{I}$ for a non trivial FI
term. Assuming that
$\sum_{I}q_{I}\delta\Psi_{I}>0$ one can see that  $m_{I}^
{2}\ge 2\sum_{I}q_{I}\delta\Psi_{I}$ always. The bound is never saturated unless $\nu=1/2, u_{i}=1, s
\sim 10^{-3}$ \footnote{I would like to thank G. Tasinato for his
  comments regarding this point.} and these values do not allow to have consistent flux
compactifications on this background compatible with tadpole
cancellation conditions, so these
other two possibilities are never achieved nor by making fine tuning
of fluxes. The unique minimum is at $C_{I}=0$.

To be sure that critical points are not saddle points if we were
interested in these cases associated to IASD fluxes, ( that we are not), we should perform the
Hessian calculation -very involved due to the higly non-trivial
structure of the $V_{T}$- , including also the F-part of the potential
that is no longer vanishing.\\

One could also ask whether  trilinear couplings of the soft terms
could change this behaviour. The answer is not for the particular case
we are considering, i.e. ISD fluxes. let see it in more detail.
The full structure of the soft and susy terms is rather complicated. Generically,
\bea
\textrm{soft
  terms}=\frac{1}{2}\sum_{a}(M_{a}\lambda^{a}\lambda^{a}+h.c.)+\sum_{I}m^{2}_{I}|C_{I}|^{2}+\\ \sum_{I,J,K}A_{IJK}C_{I}C_{J}C_{K}+1/2B_{IJ}C_{I}C_{J} + h.c.
\eea
where $\lambda^{a}$ represents gaugino mass terms, $m_{I}$ are the
scalar mass terms and there are bilinear
and trilinear couplings each of which with a highly nontrivial
dependence on all of the moduli. However one can see that all of the induced
mass that appear in the scalar potential are polynomial of lower bound $2$ in the matter fields $C_{I}$, hence, as before
$C_{I}=0$ is still an extrema, and as before $\delta\sum_{I}\Psi_{i}=0$ corresponds
to $V_{T}=0$ which is the global minimum of the theory. Moreover, the
expression for the trilinear terms in this case corresponding to
$W=W_{flux}$ is the following,
\bea
A_{IJK}= F^{M}\widehat{K_{M}}-\partial_{M}log(\widetilde{K_{II}}
\widetilde{K_{JJ}}\widetilde{K_{KK}})
\eea
trilinear terms can be estimated as
$A_{IJK}\sim 1/2(m_{3/2}-\sum_{i}1/\nu_{i})\le 1/2 m_{3/2}$ and scalar mass terms
$m_{I}^{2}\sim
m_{3/2}^{2}(1/4+3/4(-1/\sum_{i}1/\nu_{i}+\sum_{i}\frac{1}{\nu^{2}_{i}}))\ge
1/4m_{3/2}^{2}$. As before one can ask if there exists an appropiate fine
tuning in the configuration of D-branes and in the choice of 3-form
fluxes in such a way that for a particular configuration of brane case
the other two minima will be $C_{I}=0$ with a $FI \ne 0$, giving $V_{T}=0$ and the
answer is that again it seems not possible.
Both bounds are saturated for $\nu_{i}=1/2$ which
corresponds to the case cosidered above that leads to non physical
solution  in these set-ups. Scalar soft terms are the dominant
contribution. An important
clue in this result is the fact that the whole mass terms are positive
defined for ISD $G_{3}$. Otherwise there could also other solutions
corresponding to have $V_{T}<0$, and in those cases a careful analysis
should be performed.

 In the absence of magnetic fluxes moduli can not be fixed
since $\Psi_{I}$ is independent of them. However as already shown in the preceding section the
presence of magnetic fluxes does not by itself imply the
stabilization of K\"ahler moduli, as we already explained, as matter fields are free to
acquire any vev to cancel D-term contribution leaving K\"ahler
moduli unfixed. We want to emphasize that only when there are magnetic fluxes
in the 3-form flux induced soft (susy) \footnote{ This mechanism is
  also valid for supersymmetric masses generated by similar mechanisms
  whenever appropiate configurations are considered.}
breaking terms combined with the FI term, the K\"ahler moduli get fixed. This contribution
coupling to the FI-term, prevents matter fields from acquiring a vev
cancelling this contribution as it is energetically
disfavoured.

This is then a generic mechanism, in the same way that 3-form fluxes $H_{3}, F_{3}$
serve to stabilize the complex structure moduli $(2,1)$  and the
dilaton through the superpotential. The magnetic fluxes given by a particular configurations of the magnetized D-branes
fix the K\"ahler moduli (1,1) at their supersymmetric values, only once the potential has  no flat directions. These directions are lifted by the combined action of these 3-form flux-induced soft terms with magnetic fields, and the FI-terms.  We think that this solution gives a final answer to the question of perturbative stabilization
of moduli. Both types of moduli need to be present to achieve a
model without moduli. The scale of stabilization of the moduli vevs is at
supersymmetric values. The mass scale of the moduli however depends on
the model considered, and is given by the scale of the flux induced terms.

All of the previous discussion in principle remains valid for the case of susy flux induced terms, i.e. 
mass terms with the same masses for fermions and their scalar
superpartners (it has been obtained in D-brane action calculations,
see for example, \cite{camara2}, and F-theory \cite{reffert3,W}.). 
These susy mass terms, however do not appear always that there are
three form fluxes ISD (2,1) \cite{reffert2}, it is needed a D-brane
configuration with at least two chiral multiplets such that they are
at least parallel in one direction to generate $\mu$-terms. To have a
true $N=1$ model it is also needed masses for all of the moduli
present in the configuration, allowing to obtain at the same time
non-trivial areas. If such a model is provided then it
constitutes a model with K\"ahler moduli stabilized.
This implies that for a supersymmetric model, whenever the induced
flux mass terms are all positive, which is the case for ISD
fluxes, the matter fields are going to be fixed at zero. Spontaneous susy breaking will lift this value to the lowest metastable de Sitter vacua once the K\"ahler moduli have been stabilized
by the supersymmetric condition. In both cases ( non-susy and susy) this result is very appealing
since it is valid for the ISD fluxes which are the ones we are interested in since they solve the 10D equations of motion in type IIB theory. From the calculational point of view, they stabilize the K\"ahler moduli for a given Dp-brane content without the need to
specify a detailed expression for the highly complicated soft
terms. We provide an explicit example in the last section of the
paper.
\subsection{Beyond imaginary self-duality condition}

In general for IASD fluxes or a combination of both, the contribution of
$V_{F}$ has to be taken into account but this does not change the
behaviour of the stabilization.
The two possible extra extrema are
\bea 
|C_{I}|^{2}=\frac{2q_{I}\delta\Psi_{I}-m^{2}_{I}}{2q_{I}^{2}}
\eea 
For a susy model this reduces to:
\bea V(|C_{I}|^{2}_{min})&= &-\frac{\vert
m_{I}\vert^{4}}{4q_{I}^{2}}<0 \quad iff \quad m^{2}_{I}<0
\quad \forall \textsl{I}. \eea
The minima of the potential correspond to a AdS vacua.
 For the case of negative bosonic soft terms only very rough approximations to the stabilized values can
be done because of the complicated equation to solve. Moreover the
K\"ahler potential cannot be simplified so much since the bilinear $Z_{ij}$ has to
be calculated. This case  can appear when IASD fluxes are taken into
account. The explicit value at which K\"ahler moduli get fixed now it
is going to be determined by the explicit expression of the flux
induced mass terms, that one for the soft terms ( see section 4) is
highly nonlinear. On the other hand IASD fluxes do not solve the
equations of motion in this description since it is purely done at
perturbative level ( the situation changes if non-perturbative
mechanisms are taken into account), and they generate poorly
understood run-away potentials. We will not perform the calculation.

\section{An example of IIB on
$\frac{T^{6}}{Z_{2}\times Z_{2}\times \Omega R}$}
In this section we provide a concrete example of phenomenological interest which realizes the new mechanism proposed in the preceding section. Some examples of flux compactifications of phenomenological interest are \cite{mirjam3,shiu}. See also \cite{mirjam2}, in which some of the flux compactifications of phenomenological interest are able to lead to KKLT models.

In the following we will review the main features of the constructions of \cite{marchesano1, marchesano} which are a  global embedding of the local model proposed in \cite{daniel} and used in \cite{anamaria}. We will then construct our explicit realization.

The model of \cite{marchesano1,marchesano} is based on type IIB string
theory compactified on a $\frac{T^{6}}{Z_{2}\times Z_{2}}$ modded out by the
orientifold action, which has also been examined in\cite{juan}. We consider $T^{6}= \Pi_{i=1}^{3}T^{2}_{i}$. The generators of the  orbifold symmetries $Z_{2}\times Z_{2}$ are $\theta,\omega$ which act on the complex coordinates of the tori as
\bea
\theta : (z_{1},z_{2},z_{3})\to (-z_{1},-z_{2},z_{3}),\\
\omega : (z_{1},z_{2},z_{3})\to (z_{1},-z_{2},-z_{3}).
\eea
The orientifold action is given by $\Omega R$ where $\Omega$ is the
usual world-sheet parity and \\
$R:(z_{1},z_{2},z_{3})\to (-z_{1},-z_{2},-z_{3}).$ This model contains
64 $O3$ planes each one on a fixed point of $\mathcal{R}$ and 4
$O7_{i}$ planes, located at the $Z_{2}$ fixed points of the i-th
$T^{2}$ and wrapping the other tori. We consider the case of intrinsic
torsion as in \cite{shiu}. The open string sector contains D9 branes
with non trivial magnetic fluxes. The non-trivial gauge bundle
generically reduces the rank of the group and upon KK reduction leads
to D=4 chiral fermions. The magnetic fluxes also induce D-brane
charges of lower dimension that contribute to the tadpoles. Magnetized
D9 branes usually have D7, D5 and D3 charges.
As explained in \cite{juan} the topological information of these models
is encoded in three numbers. $(N_{a},(n_{a}^{i},m_{a}^{i}))$ where
$N_{a}$ is the number of D9 branes contained on the $a$-stack,
$m_{a}^{i}$ is the number of times that a-stack of D-branes wrap
the i-th $T^{2}$ and $n_{a}^{i}$ is the units of magnetic flux in that
torus induced by D-branes. The unit of magnetic flux of the D-branes
in that torus, as seen in section 3, is

\bea
\frac{m_{a}^{i}}{2\pi}\int_{T_{i}^{2}}F_{a}^{i}=n_{a}^{i}.
\eea

The $D9_{a}$ branes preserve the same
supersymmetry of the orientifold planes provided that \bea
\sum_{i=1}^{3}\Psi_{a}^{i}&=& \frac{3\pi}{2}\quad mod(2\pi),\\
\textrm{with}\quad\pi\Psi_{a}^{i}&=&arctan{(\frac{n_{a}^{i}\beta_{i}}{m_{a}^{i}})},
\eea
where no summation over index $i$ is performed.
This condition also guarantees that any two sets of branes
preserve a common supersymmetry since the relative angles \bea
\theta_{ab}^{i}=\Psi_{b}^{i}-\Psi_{a}^{i} \eea  trivially satisfy
\bea
\sum_{i}\theta_{ab}^{i}=0\quad mod (2\pi).
\eea
In the case of a global analysis we need to add new branes which are
expected to be in a hidden sector in order to cancel global tadpoles \cite{juan}. The
conditions are,
\bea
& 1)&\sum_{\alpha}N_{\alpha}n_{\alpha}^{1}n_{\alpha}^{2}n_{\alpha}^{3}+1/2N_{fluxes}=16\quad
N_{min}=8\quad \textrm{with torsion}\\
&
2)&\sum_{\alpha}N_{\alpha}m_{\alpha}^{i}n_{\alpha}^{j}m_{\alpha}^{k}=-16\quad
i\ne j\ne k \quad\textrm{and}\quad i,j,k=1,2,3 \eea
 with $N_{flux}=64.n\quad n\in Z$.
Global cancellation of ${\bf Z_{2}}$ RR charges must also be imposed
to cancel the contribution of $D5_{i}-\overline{D5_{i}}$ and
$D9_{i}-\overline{D9_{i}}$ pairs, which is equivalent to satisfying the
following conditions for a case with torsion,
\bea
&&\sum_{\alpha}N_{\alpha}m_{\alpha}^{1}m_{\alpha}^{2}m_{\alpha}^{3}\in
{\bf 4Z}\\
&&\sum_{\alpha}N_{\alpha}n_{\alpha}^{i}n_{\alpha}^{j}m_{\alpha}^{k}\in
{\bf 4Z}\quad i\ne j\ne k \quad\textrm{and}\quad i,j,k=1,2,3
\eea
and for the case without torsion the conditions are the same by making these changes,
\bea
&& \sum_{\alpha}N_{\alpha}n_{\alpha}^{1}n_{\alpha}^{2}n_{\alpha}^{3}+1/2N_{fluxes}=-16\quad
N_{min}=4\\
&&\sum_{\alpha}N_{\alpha}m_{\alpha}^{1}m_{\alpha}^{2}m_{\alpha}^{3}\in
{\bf 8Z}\\
&&\sum_{\alpha}N_{\alpha}n_{\alpha}^{i}n_{\alpha}^{j}m_{\alpha}^{k}\in
{\bf 8Z}\quad i\ne j\ne k \quad\textrm{and}\quad i,j,k=1,2,3 \eea
Let us focus on the case with
torsion. The following model is a concrete realization that serves
our purpose.
\subsubsection*{MSSM}
\bea
\begin{tabular}{|l|c|c|c|c|}
\hline & & & & \\
${\bf N_{\alpha}}$ & $(n^{1}_{\alpha},m^{1}_{\alpha})$ & $(n^{2}_{\alpha},m^{2}_{\alpha})$ &
$(n^{3}_{\alpha},m^{3}_{\alpha})$ &
$(\Psi^{1}_{\alpha},\Psi^{2}_{\alpha},\Psi^{3}_{\alpha})$\\ & & & & \\
\hline ${N_{a}= 8}$ & (1,0) & (g,1) & (g,-1) & $(
\frac{\pi}{2},\pi\delta_{2},\pi-\pi\delta_{3})$\\
${ N_{b}= 2}$ & (0,1) & (1,0) & (0,-1) & $(0,\frac{\pi}{2},\pi)$\\
${ N_{c}= 2}$ & (0,1) & (0,-1) & (1,0) & $(0,\pi,\frac{\pi}{2})$
\\ \hline ${ N_{h_{1}}= 2}$ & (-4,1) & (-2,1) & (-3,1) &
$(\pi(1-\varphi_{1}),\pi(1-\varphi_{2}),\pi(1-\varphi_{3}))$\\
${N_{h_{2}}=2}$ & (0,1) & (-5,1) & (-4,1) &
$(0,\pi(1-\phi_{2}),\pi(1-\phi_{3}))$\\
\hline ${ 8N_{f}}$ & (1,0) & (1,0) & (1,0) &
$(\frac{\pi}{2},\frac{\pi}{2},\frac{\pi}{2})$\\
\hline
\end{tabular}
\eea
\\
This includes a slight modification to the proposal of
\cite{shiu} since their configuration was not able to fix all of the
K\"ahler moduli. In fact the spectrum keeps its interesting
phenomenological properties except for
the number of families which depend on the particular homological
charges chosen. To satisfy the consistency conditions the following
equation has to be solved,
\bea
4n+g^{2}+N_{f}=8. \eea It gives different flux vacua for the
different possible replication of families $I_{ab}=\Pi_{i}(n_{a}^{i}m_{b}^{i}-n_{b}^{i}m_{a}^{i})$. Searching for a
realistic scenario it is necessary to obtain three family
replication, unfortunately a simple examination reveals that this
model does not contain it. The possible vacua are:\\
For $g=2$ \bea n &=& 0 \quad N_{f}=4\\ n &= &1 \quad N_{f}=0 \eea
For $g=1$ \bea n &=& 0 \quad N_{f}=7\\ n &= &1 \quad N_{f}=3 \eea
To illustrate our example we choose the vacua for $g=1$, containing
fluxes, i.e.
$n=1, N_{f}=3$.
\subsubsection*{Spectrum}
Locally the spectrum of this model can contain the MSSM since $g$ is
not constrained. This has been studied in \cite{daniel}. This sector
is called the visible sector. We show it below as a remainder,

\bea
\begin{tabular}{|c|c|c|c|c|}
\hline Intersection & Matter fields & Rep & $Q_{B-L}$& Y\\ \hline
$D7_{1}-D7_{2}$ & $Q_{L}$ & $ 3(3,2)$ & 1 & $1/6$\\ \hline
$D7_{1}-D7_{3}$ & $U_{R}$ & $ 3(\overline{3},1)$ & -1 & $-2/3$\\
\hline $D7_{1}-D7_{3}$ & $D_{R}$ & $ 3(\overline{3},1)$ & -1 &
$1/3$\\ \hline $D7_{1}-D7_{2}$ & $E_{L}$ & $ 3(1,2)$ & -1 &
$1/2$\\ \hline $D7_{1}-D7_{3}$ & $E_{R}$ & $ 3(1,1)$ & 1 & $-1$\\
\hline $D7_{1}-D7_{3}$ & $N_{R}$ & $ 3(1,1)$ & 1 & 0\\ \hline
$D7_{2}-D7_{3}$ & H & $ (1,2)$ & 0 & $1/2$\\ \hline
$D7_{2}-D7_{3}$ & $\overline{H}$& $(1,2)$ & 0 & $-1/2$\\ \hline
\end{tabular}
\eea
Table 1: Chiral spectrum of the MSSM-like model.
\\

However the global completion imposes constraints in such a way that
$g\ne 3$ so it is not possible 
to obtain 3-family replication and in that sense,  
we are proposing is a toy model although it keeps the nice properties of chiral matter, correct quantum numbers, etc..
It contains the intersection of the so-called, visible sector and the hidden
sector and hidden-hidden sector.  We analyze the full spectrum in terms of a Pati-Salam model.
\\
\vspace{1cm}
\\
\bea
\begin{tabular}{|c|c|c|c|c|c|c|}
\hline & & & & & & \\ Sector & Matter & SU(4)xSU(2)xSU(2)xUSp[24]&
$Q_{a}$ & $Q_{h_{1}}$ & $Q_{h_{2}}$ & $Q^{'}$\\ \hline $(ab)$ &
$F_{L}$& $(4,2,1)$ & 1 & 0 & 0 & 1/3
\\ \hline (ac)& $F_{R}$ &  ($\overline{4}$,1,2) & -1 &0 &0& -1/3\\
\hline (bc)& H & (1,2,2) & 0 &0&0&0\\ \hline $(ah_{1})$ & &
$6(\overline{4},1,1)$ & -1 & -1 &0 &5/3\\ \hline $(ah_{1}^{'})$ &
& $4(\overline{4},1,1)$ & -1 & -1 &0 &5/3\\ \hline $(ah_{2})$ & &
18(4,1,1)& 1 & 0 & -1 & -5/3\\ \hline $(ah_{2}^{'})$ & &
20(4,1,1)& 1 & 0 & -1 & -5/3\\ \hline $(bh_{1})$ & & 12(1,2,1)& 0&
-1 & 0 & 2\\ \hline $(bh_{1}^{'})$ & & 12(1,2,1)& 0 & -1 & 0& 2\\
\hline $(ch_{1})$ & & 8(1,1,2)& 0& -1 & 0 & 2\\ \hline
$(ch_{1}^{'})$ & & 8(1,1,2)& 0& -1 & 0 & 2\\ \hline
$(h_{1}h_{1}^{'})$& & 288(1,1,1)& 0& -2 & 0 & 4\\ \hline
$(h_{1}h_{2})$ & & 12(1,1,1)& 0& 1 & 1 &0\\ \hline
$(h_{1}h_{2}^{'})$ & & 196(1,1,1)& 0& 1 & 1 &0\\ \hline $(fh_{1})$
& & (1,1,1)x[24]& 0 &-1 &0&2\\ \hline $(fh_{2})$ & &
(1,1,1)x[24]&0&0& -1&2\\ \hline
\end{tabular}
\eea
Table: Chiral spectrum of a one generation Pati-Salam $N=1$ chiral
model of table 1. The abelian generator of the unique massless
U(1) is given by $Q^{'}=\frac{1}{3}Q_{a}-2(Q_{h_{1}}-Q_{h_{2}})$.
\\

Some linear combinations of U(1) will leave $U(1)$ fields massive in
a Green-Schwarz mechanism but since our purpose is to show the
perturbative stabilization of moduli we have not included them in the
spectrum calculation. 
We show that a realization of string compactification on a $CY_{3}$
orientifold with some phenomenological properties render the moduli
fixed.

When we substitute for the supersymmetry conditions,by taking the
values of table $(6.15)$ in  $(6.4)$, we obtain the
following equations 
\bea && \delta_{2}=\delta_{3}\to t_{2}=t_{3},\\ && \pi(\varphi_{1}+\varphi_{2}+\varphi_{3})=\frac{3\pi}{2},\\ &&
\pi(\phi_{2}+\phi_{3})=\frac{\pi}{2}. \eea 
Using $(6.5)$ a straightforward
calculation gives \bea \beta_{1}=0.157,\quad
\beta_{2}=\beta_{3}=\frac{1}{\sqrt{20}} \eea which means that the
vacuum expectation value at which K\"ahler moduli gets fixed is,
\bea t_{1}=20s,\quad t_{2}=t_{3}=127.4s, \eea where
$s$ denotes the vacuum expectation value of the dilaton which is the
string coupling constant. Substituting in the value of areas ($3.6$) this
means that \bea
\begin{tabular}{|c|c|}
\hline & \\ $A_{1}=6.38\alpha^{'}$ &
$A_{2}=A_{3}=4.47\alpha^{'}$\\
\hline
\end{tabular}
\eea
\\
This mechanism implies that for this toroidal compactification, are
needed three and only three stacks of
D-branes with different magnetic fluxes in
 such a way that lead to different equations in order to have a
 determined compatible system of equations. This is the subtlety that
 does not allow us  to use Marchesano et al.'s model in our mechanism
 as an example of D-brane configuration since they have just two different
equations. This construction is in no way unique, and represents a toy
model to illustrate the mechanism of stabilization. We expect that
more complicated models can be constructed with realistic spectrum and
with all of the moduli fixed.

Regarding the blow-up moduli associated to the orbifold fixed points,
they are  going to be also stabilized with this mechanism since for models
with intrinsic torsion, the moduli are K\"ahler, and can get fixed in the
same way  as it was done in \cite{throat}. 
With the above  results we can perform the explicit
calculation of soft mass terms for a given three-form flux
configuration with complex-structure moduli and dilaton
stabilized. We choose the one \cite{kachru1} for $n=1,
N_{flux}=64$ with $\mathcal{N}=0$ supersymmetry \bea
G_{3}&=&2(d\overline{z}_{1}dz_{2}dz_{3}+dz_{1}d\overline{z}_{2}dz_{3}+
dz_{1}dz_{2}d\overline{z}_{3}+d\overline{z}_{1}d\overline{z}_{2}d\overline{z}_{3}),\\
W &=& 8(u_{1}u_{2}u_{3}- s).
\eea The complex-structure moduli and the dilaton are stabilized
at $u_{i}=s=i$, which in our notation is equivalent to $u_{i}=s=1$. This
values lead to a non-acceptable value
for perturbative analysis since $g_{s}=1$ as explained in
\cite{marchesano}. However we will use this to give an example of explicit calculation of the
soft terms. The soft terms from T-dominance (ISD) fluxes
correspond to $V_{0}=0$. The gravitino vev is \bea
m_{3/2}^{2}=\frac{|W|^{2}}{s\Pi_{i}t_{i}u_{i}}\eea and we have
made $C=1, cos(\theta)=1, \eta_{i}=1/\sqrt{3},\gamma_{i}=\gamma_{T}$
as is explained in detailed in \cite{anamaria}.

The soft breaking terms for this model are: \bea
\begin{tabular}{|c|c|c|}
\hline & & \\ $m_{ab}^{2}=375,2 m_{3/2}^{2}$ & $m_{ac}^{2}=7,8
m_{3/2}^{2}$ & $m_{bc}^{2}=\frac{1}{4} m_{3/2}^{2}$\\ \hline & &\\
$m_{ah_{1}}^{2}=120,2 m_{3/2}^{2}$ & $m_{ah_{2}}^{2}=58,5
m_{3/2}^{2}$ & $m_{af}^{2}=16,9 m_{3/2}^{2}$\\
\hline & &\\ $m_{bh_{1}}^{2}=79,9 m_{3/2}^{2}$ &
$m_{bh_{2}}^{2}=222,3 m_{3/2}^{2}$ & $m_{bf}^{2}=\frac{1}{4}
m_{3/2}^{2}$\\ \hline & & \\ $m_{ch_{1}}^{2}=125,6 m_{3/2}^{2}$ &
$m_{ch_{2}}^{2}=225,2 m_{3/2}^{2}$ & $m_{cf}^{2}=\frac{1}{4}
m_{3/2}^{2}$\\ \hline & & \\ $m_{h_{1}h_{2}}^{2}=244,8
m_{3/2}^{2}$ & $m_{h_{1}f}^{2}=39,4 m_{3/2}^{2}$ &
$m_{h_{2}f}^{2}=30,5 m_{3/2}^{2}$\\ \hline
\end{tabular}
\eea in terms of gravitino mass $m_{3/2}$. Since the
microscopic source of SUSY-breaking is the above ISD flux in a toroidal setting, by using the above definitions, $|W|^{2}=256$, and the gravitino mass is
$m_{3/2}=0.019$, so the soft terms have unrealistic values as they are extremely high.

For general toroidal/orbifold models with intersecting D6-branes
the flux-induced soft terms are typically of the order of the string
scale $(1/\alpha^{'})$, which is only  slightly smaller than $M_{Pl}$
and not able to solve hierarchy problems \cite{anamaria}. This fact is due to the simplicity of the compactification
manifolds as well as the fact that fluxes are distributed uniformly.

\section{Discussion and conclusions}
We have shown a new method to dynamically stabilize, with
fluxes, all moduli in supersymmetric and non supersymmetric models.
In supersymmetric models we have explained that the mechanism is
restricted to particular configurations able to generate $\mu$ terms
for all of the moduli, i.e. configurations of stacks parallel to each
other in at least one direction in which a suitable $(2,1)$ $G_{3}$
flux has been turned on. 
Three-form fluxes generically stabilize the dilaton and the
complex structure moduli. They generate an F-term part of the scalar
potential. Magnetic fluxes are two forms that fix K\"ahler moduli at
their supersymmetric value once the potential has no flat
directions. This goal is achieved by inducing through 3-form flux
a non-susy (susy) breaking (flux-induced) mass terms with magnetic fields in
the scalar potential, that combined with the FI term lift
the flat directions. We think this is a generic mechanism that
gives a final answer to the problem of perturbative moduli
stabilization. These mass terms avoid the possibility of cancelling the D-term
by the consistent adjustment of the matter fields vev, since it requires an energy cost. To leading order, the F-part of the
potential is a non-scale potential and together with D-term are
responsible for the stabilization of the K\"ahler moduli. ISD
fluxes induce positive mass terms and
fix the K\"ahler value of the D-term to its supersymmetric value. In
the supersymmetric case it can be possible to implement a similar
mechanism of the one found in \cite{quevedo} to generate a de Sitter space (by
spontaneous supersymmetry breaking ). However,
possibly toroidal models will be unable to generate adequate
fine tuning to guarantee that the different approximations are
still valid (small cosmological constant, small quantum
corrections, a big potential barrier). Maybe this mechanism
combined with the one of \cite{quevedo} could be implemented in a more
complicated model (it is necessary that $U(1)$'s of the FI term will
not be charged or the matter will have some special properties
that are not present in the case considered). For the
case $m^{2}<0$
both parts of the potential, F- and D-, including the supergravity
potential $V_ {F}^ {background}$
contribute and an explicit calculation in terms of the particular soft breaking terms is needed. 
They can be associated to IASD or a combination of (IASD and ISD)
fluxes. The analysis of IASD contribution has not been performed as
they do not lead to solutions of physical interest. The scalar potential generically has two AdS minima.

Clearly the case with $m^{2}>0$ which correspond to turning on ISD
three form fluxes is much more interesting since it solves the
equation of motion and stabilizes the moduli at a value of the
order of the string scale which means that it does not depend on the scale
of supersymmetry breaking and is higher enough to induce
reheating processes at early stages of the universes. The mass
scale of the moduli presumably is of the order of the soft
breaking mass scale (which is of the order of flux scale
$\frac{\alpha^{'}}{R^{3}}$). We
have given a concrete  K\"ahler moduli-free realization of phenomenological interest of Type IIB on
$\frac{T^{6}}{Z_{2}\times Z_{2}\times \Omega R}$. However this model
represents a toy model. We expect improvements in the search for
realistic compactifications moduli-free (i.e. three family generation,
lower soft masses) in more complicated  scenarios
as those with warped metrics due to throats, that are also able to explain
the hierarchy problem, as well as, address inflation.

\section{Acknowledgements}
I would like to thank to the referee for his/her comments that have
helped me very much to improve the paper.
I want to thank J.L.F.Barb\'{o}n, D. Cremades, J. Conlon,
E. L\'{o}pez, L. Iba\~{n}ez, F.Quevedo and G. Tasinato for useful conversations. I am
also very grateful  to A. Font for many comments and clarifying explanations about her work. Finally, I am specially indebted
  to Angel Uranga for his continuous orientation and discussions all throughout this work.
 M.P.G.M. is supported by a postdoctoral 
grant of the Consejer\'{\i}a de Educaci\'on, Cultura, Juventud y 
Deportes de la Comunidad Aut\'onoma de La Rioja (Spain).



\end{document}